\documentclass[12pt]{article} 
\usepackage{epsfig}
\usepackage[usenames,dvips]{color}
\definecolor{darkgreen2}{rgb}{0.0,0.6,0.0}
\definecolor{darkgreen}{gray}{0.0}

\begin{document}
\centerline{\Large \bf  Muon Emittance Exchange with a Potato Slicer}

\bigskip

\centerline{\large D.\,J.\,Summers, T.\,L.\,Hart, J.\,G.\,Acosta, L.\,M.\,Cremaldi,}
\smallskip
\centerline{\large S.\,J.\,Oliveros,  and L.\,P.\,Perera}
\smallskip
\centerline{\large University of Mississippi\,-\,Oxford, University, MS 38677 \ USA}
\bigskip

\centerline{\large D.\,V.\,Neuffer}
\smallskip
\centerline{\large Fermilab, Batavia, IL 60510 \ USA}
\medskip

\centerline{MAP-DOC\,-\,4405, \ Apr 2015}

\bigskip
\centerline{\large \bf Abstract}
\medskip

We propose a novel scheme for final muon ionization cooling with quadrupole doublets followed by emittance exchange in vacuum
to achieve the small beam sizes needed by a muon collider.
A flat muon beam with a series of quadrupole doublet half cells appears to provide the strong focusing required for final cooling.
Each quadrupole doublet has a low beta region occupied by  a dense, low Z absorber.
After final cooling,
normalized transverse, longitudinal, and angular momentum
emittances of 0.100, 2.5, and 0.200 mm-rad are exchanged into
0.025, 70, and 0.0 mm-rad.
A skew quadrupole triplet transforms a round muon bunch with modest
angular momentum into a flat bunch with no angular momentum. Thin
electrostatic septa efficiently slice the flat bunch
into 17 parts.
The 17 bunches are interleaved into a 3.7 meter long train with RF deflector cavities.
Snap bunch coalescence combines the muon
bunch train longitudinally in a 21 GeV ring in 55 $\mu$s, one quarter of a synchrotron oscillation period.
A linear long
wavelength RF bucket gives each bunch a different energy causing the
bunches to drift in the ring until they merge into one bunch and can be
captured in a short wavelength RF bucket with  a 13\% muon decay loss and a packing fraction as high as 87\,\%.


\begin{table}[b!]
   \caption{Helical and  Rectilinear  Cooling Channel    normalized  6D emittances from simulations and the normalized 6D emittance needed for
muon collider. The channels cool by over five orders of magnitude and need less than a factor of 10 more for a  muon collider.
The 21 bunches present after phase rotation are also merged into one bunch during the 6D cooling.}
\vspace{-2mm}
\tabcolsep = 2.5mm
\begin{center}
\renewcommand{\arraystretch}{1.20}
 \begin{tabular}{lccccc}
   & $\epsilon_x$\,(mm) & $\epsilon_y$\,(mm) & $\epsilon_z$\,(mm)  & $\epsilon_{6D}$\,(mm$^3$) & Ref. \\ \hline
 Emittance after Phase Rotation & 48.6 & 48.6 & 17.0  & 40,200 & \cite{Stratakis}\\  
 Helical Cooling Channel & 0.523 & 0.523 & 1.54 &   0.421 & \cite{Yoshikawa} \\
 Rectlinear Cooling Channel & 0.28 & 0.28 & 1.57 &   0.123 & \cite{Stratakis}\\
 Muon Collider & 0.025 & 0.025 & 70 &  0.044 & \cite{Neuffer94} \\ \hline
 \end{tabular}
\end{center}
\label{table:channels}
\end{table}

\begin{table}[t!]
   \caption{Rectilinear cooling channel final cell beta function\,\cite{Stratakis2}. p = 204 MeV/c, $\beta \, \gamma$ = p/mc = 204/105.7 = 1.93,  $\epsilon_{x,y}^N$ = 0.280\,mm,
 $\sigma_{x,y} = \sqrt{\beta_{x,y} \, \epsilon_{x,y}^N / (\beta \, \gamma)}$, and     
  $\theta_{x,y} = \sqrt{\epsilon_{x,y}^N / ( \beta_{x,y} \, \beta \, \gamma)}$.
 A bore diameter of 8 x 7.86\,mm = 62.9\,mm contains $\pm 4 \,\sigma_{x,y}$ when  $\beta_{\,x,y}$ = 42.64\,cm. The regions where $\beta$ is near 3\,cm are short.}
\vspace{-2mm}
\tabcolsep = 1.7mm
\begin{center}
\renewcommand{\arraystretch}{1.20}
 \begin{tabular}{lccccccccccc} \hline
z(m)                & 0.000  & 0.016  & 0.030 & 0.092 & 0.183  & 0.402 & 0.625 & 0.717 & 0.779 & 0.793 & 0.806 \\
$\beta_{\,x,y}$(cm)   & 3.08    & 3.93    &  5.92   & 26.16 &42.64  & 33.75 & 42.64 & 26.16 & 5.92    & 3.93      & 3.10\\ 
$\sigma_{x,y}$(mm) & 2.11 & 2.39 & 2.93 & 6.16 & 7.86 & 7.00 & 7.86 & 6.16 & 2.93 & 2.39 & 2.12 \\
$\theta_{x,y}$(mrad) & 68.6 & 60.7 & 49.5 & 23.5 & 18.4 & 20.7 & 18.4 & 23.5 & 49.5 & 60.7 & 68.4 \\ \hline 

\end{tabular}
\end{center}
\label{table:Rectilinera-beta}
\end{table}

\begin{table}[b!]
 \caption{Muon equilibrium emittance at 200 MeV/c (KE = 121 MeV, $\beta$ = v/c = 0.88) for hydrogen gas, lithium hydride, beryllium, boron carbide,  diamond, and beryllium oxide\,\cite{Neuffer2013, PDG2014}.  
$\epsilon_{\perp} = \beta^* E_s^{\,2} / (2 g_x \, \beta \, m_{\mu} c^{\,2} (dE/ds) L_R)$,
 where $\beta^*$ twiss is  1\,cm,  $E_s$ is 13.6 MeV,    the transverse damping partition number, $g_x$ is one with parallel absorber faces,
$m_{\mu} c^{\,2}$ is 105.7 MeV, and L$_R$ is radiation length. 
}   
\vspace{-2mm}
\tabcolsep = 2.5mm
\begin{center}
\renewcommand{\arraystretch}{1.20}
 \begin{tabular}{ccccc} \hline
Material          &     Density        & L$_R$ & dE/ds         & $\epsilon_{\perp}$\,(equilibrium) \\ 
                         &    g/cm$^3$   &      cm          & MeV/cm    & mm\,-\,rad\\ \hline
H$_2$ gas & 0.000084       & 750,000    &  0.00037  & 0.036 \\
Li\,H       &  0.82           & 97    & 1.73  & 0.059 \\
Be        & 1.85         & 35.3   & 3.24  & 0.087 \\ 
B$_4$C & 2.52       & 19.9   & 4.57 & 0.109 \\
Diamond & 3.52     &   12.1 &  6.70         &   0.123         \\ 
Be\,O   & 3.01         &  13.7  &  5.51         &  0.132          \\ \hline
\end{tabular}
\end{center}
\label{table:equilibrium}
\end{table}

\bigskip
\bigskip
\leftline{\large \bf Introduction}
\medskip

Due to s-channel production, a muon collider\,\cite{Neuffer94} may be ideal for the examination of   H/A Higgs scalars which could be at the 1.5 TeV/c$^{\,2}$ mass scale and are 
required in supersymmetric models\,\cite{Eichten}.
But what is the status of  muon cooling?
As noted in Table~1, more than  five orders of magnitude of muon cooling have been shown in two simulated designs\,\cite{Yoshikawa,Stratakis} 
but not quite the six orders of magnitude needed
for a high luminosity muon collider.  Also as noted in Table 1,  some of the longitudinal cooling needs to be exchanged for lower transverse emittance. 

The breakdown of RF cavities operating in strong magnetic fields is an issue\,\cite{Moretti}.  
The Helical Cooling Channel inhibits breakdown with high pressure hydrogen\,\cite{Hydrogen}.
Hydrogen at moderate pressures, lower than those used in interstate natural gas pipelines, may work for the Rectilinear Cooling Channel\,\cite{Hydrogen2}. 
As seen in Table 2, the Rectlinear Cooling Channel does have some high $\beta$ regions, but these only cause minimal heating if hydrogen pressure is modest\,\cite{Hydrogen3}. 

An infinite solenoid\,\cite{Gallardo} with a 14 Tesla magnetic field and  a 200 MeV/c muon beam gives a betatron function of   
$ \beta_{\perp} = 2 \, p / (3.0 B) = 2 (200 \, {\rm{MeV/c}})/[ 3.0 \,(14{\,\rm{ T}})]$ = 9.5\,cm. As noted in Table 2, the short solenoids in the final stage of the 
Rectilinear Cooling Channel give a betatron function of 3.1\,cm which is used with lithium hydride.  To possibly get to 
the lower betatron values of about 1\,cm needed by 
a muon collider for final cooling\,\cite{final}, quadrupole doublet cells are explored in Appendix Z. Table 3 gives transverse equilibrium  emittances for a number of low Z materials,
particularly those with high densities.

\bigskip
\leftline{\large \bf Round to Flat Beam Transformation}

\begin{figure}[b!]
\begin{center}
\epsfig{figure=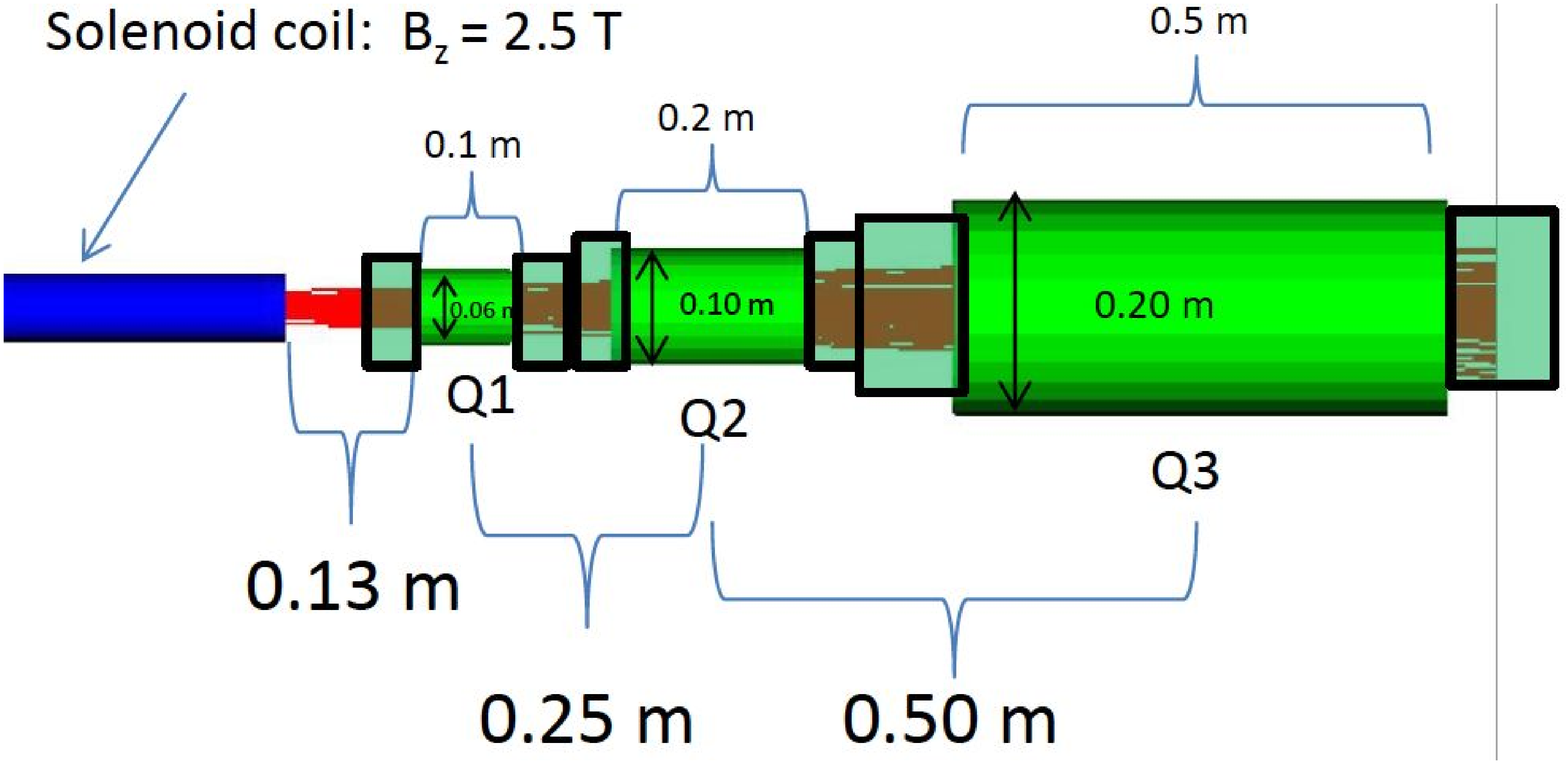,height=45mm} \
\epsfig{figure=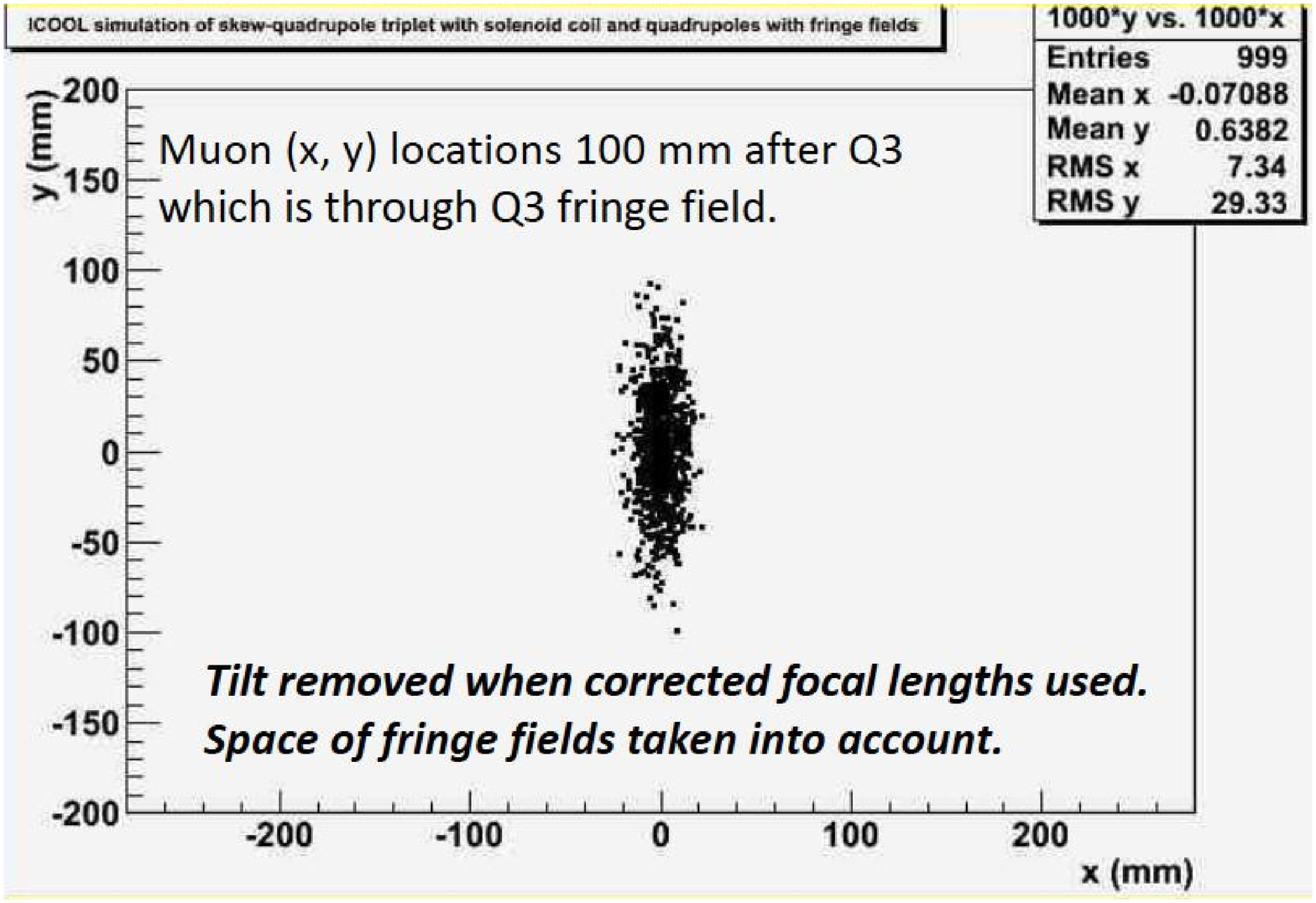,height=45mm}
\caption{A round, spinning 200$\,\pm$10 MeV/c muon beam with a normalized transverse emittance of 0.100\,mm is transformed to a flat, non-spinning muon beam with new normalized
transverse emittances of 0.0264\,mm and 0.4206\,mm\,\cite{Brinkmann}. 
Inside the 2.5\,T solenoid $\sigma_{x,y}$~=~7.5\,mm.
The skew quadrupole pole tip fields and focal lengths are (0.77, 0.45, 0.12)T and (0.259, -0.369, 1.073)m, respectively.
The calculated thin quadrupole parameters must be adjusted somewhat  to make finite length quadrupoles work properly.
Fringe fields are included.
The simulation was
done with ICOOL\,\cite{Fernow2005}.}
\label{figure:Round}
\end{center}
\end{figure}

Assume that a muon beam with a normalized transverse emittance of 0.100\,mm is available.
Further assume that  the beam has some modest and smooth residual angular momentum coming out of a
6D cooling channel.  
The beam does pick up canonical angular momentum as it passes though absorbers in solenoids.

First, a round spinning muon beam with angular momentum 
is transformed  to a flat non-spinning beam with a skew quadrupole triplet\,\cite{Brinkmann} as shown in Fig.\,1.  This should make slicing easier.  
The $x$ to $y$ emittance ratio of the flat beam is 
$(\sqrt{\epsilon^2 + L^2} + L) / (\sqrt{\epsilon^2 + L^2} - L)$,
where $\epsilon$ is the intrinsic normalized transverse emittance,
$L= eB\,\sigma_{x,y}^{\,2}\,/2mc$ is the  quadrature emittance contribution of the canonical angular momentum, $B$ is the solenoidal field strength that the beam is exiting, and 
$\sigma_{x,y}$ is the round beam radius. 
$L$ is chosen to give an emittance ratio of 16 for a beam with new nominal transverse normalized  emittances of 0.0264\,mm and 0.4206\,mm.  The muon momentum is 
200$\,\pm\,$10 MeV/c.  Higher momentum spreads dilute performance, but beam preconditioning may help\,\cite{Zhu}.

\bigskip
\leftline{\large \bf Slice the Flat Beam with 16 Septa}
\medskip




\begin{table}[b!]
 \caption{Combine 17 bunches into a 3.7\,m long train  with 10 RF Deflector Cavities. Each cavity interleaves two or three bunch trains.
 Deflection is $\pm$4.5\,mrad or zero at 300 MeV/c.
 The final train has a 231\,mm bunch spacing for acceleration by 1300\,MHz RF cavities.}   
\vspace{-2mm}
\tabcolsep = 1.4mm
\begin{center}
\renewcommand{\arraystretch}{1.20}
 \begin{tabular}{ccccccc} \hline
 &  Number & Number of & RF   & RF   & Output  & Output \\
 Tier &   of Trains & RF Deflector &  Frequency  & Wavelength         & Spacing in & Bunch     \\ 
   & Interleaving & Cavities & MHz   &     &   Wavelengths & Spacing    \\ \hline
1 & 17 $\to$ 6 & 6 &  (9/16)1300 = 731  & 410\,mm        & 9/4      & 923\,mm  \\
 2 & 6 $\to$ 2&  3 & (3/8)1300 = 487 & 616\,mm   & 3/4       & 462\,mm  \\
3 & 2 $\to$ 1 & 1 & (1/2)1300 = 650     & 462\,mm    & 1/2      & 231\,mm      \\ \hline
 \end{tabular}
\end{center}
\label{table:deflector}
\end{table}

\medskip
Slice\,\cite{Syphers}  the flat muon beam into 17 pieces.  
The slice width is chosen to give a horizontal emittance of 0.025 mm-rad and includes 91\% of the muons;  9\% of the muons are in the tails and lost.  
A $w = $ 0.1\,mm wide electrostatic septa was used with 98\% efficiency at the Fermilab Tevatron fixed target program for
multiturn extraction. At the Tevatron the fractional loss was given by

\begin{equation}
 {4 \, \sqrt{2} \, \, w \over x_{\rm{max}} \sqrt{\beta_s/\beta_0}} = {4\,\sqrt{2} \times 0.1\,{\rm{mm}} \over 20\,{\rm{mm}} \sqrt{2.3}} = 0.02,
\end{equation}

\noindent
where $x_{\rm{max}}$ is the beam size with the usual $\beta_{\,0}$ of the lattice and a larger $\beta_{\perp}$ function, $\beta_s$, is used in the extraction region to make the beam bigger.
The physical width of the muon beam is 
 
 \begin{equation}
 \sqrt{\epsilon_{N,x} \, \beta_x \over \beta\,\gamma} = \sqrt{{(0.4206\,{\rm{mm}})(1,000\,{\rm{mm})} \over (0.93)(2.72)}} = 13\,{\rm{mm}}, 
 \end{equation}

\noindent 
giving a modest 3\% slicing loss.  
The geometrical width of the beam needs to be much larger than the width of the electrostatic septa for efficient slicing.

\bigskip
\leftline{\large \bf Create a 3.7\,m long bunch train with RF deflector cavities}
\medskip

Combine 17 bunches into a 3.7 m long train with RF deflector cavities as used in CLIC tests. 
Each cavity interleaves two or three bunch trains. Deflection is ±4.5 mrad or zero at 300 MeV/c\,\cite{CLIC}.
The final train has a 231 mm bunch spacing for acceleration by 1300 MHz RF cavities (see Table 4).
Estimate the required kick\,\cite{Neuffer-Injection}  to inject a  
300 MeV/c ($\gamma \beta$ = 300/105.7 = 2.84) beam with a normalized emittance of
0.025 mm-rad and a $\beta^*$ of 8000\,mm.
The kick must be 4x greater than the rms divergence of the beam or 
4 $\sqrt{\epsilon/(\gamma\beta\beta^*)} $ = 4.2 mrad, which matches CLIC.
The $\pm \, 4 \sigma$ beam diameter is 
8$\sqrt{\epsilon  \beta^* /(\gamma\beta)} $ = 67\,mm.

\bigskip
\leftline{\large \bf Snap bunch coalesce a train of 17 bunch into one in a ring}
\medskip

\medskip
Finally,  {\it snap bunch coalescing} with RF  is used to combine the 17 muon bunches longitudinally.  
In snap bunch coalescing, all bunches are partially rotated over a quarter of a synchrotron period in energy-time space with a linear long wavelength 
RF bucket and then the bunches drift in a ring until they merge into one bunch and can be captured in a short wavelength RF bucket.
The bunches drift together because they each have a different energy set to cause the drift.
Snap bunch coalescing replaced adiabatic bunch coalescing at the
Fermilab Tevatron collider program and was used for many years\,\cite{Foster}. Sets of fifteen bunches were combined in  the Tevatron.  A 21~GeV ring
has been used in a simulation\,\cite{Johnson} with ESME\,\cite{ESME} to show the coalescing of 17  muon bunches in  55~$\mu$s.
The lattice has $\gamma_{\,t}$ = 5.6\,\cite{Bogacz}.
The muon decay loss  is 13\%. The longitudinal packing fraction is as high as 87\%\,\cite{Bhat}.  So nominally, the initial normalized 2.5\,mm
longitudinal emittance is increased by a
factor of 17/0.87 to become 49\,mm, which is less than the 70\,mm needed for a muon collider.


\bigskip
\leftline{\large \bf Conclusions}
\medskip

Emittance exchange with a potato slicer may  be able to achieve the final normalized 0.025\,mm transverse and 70\,mm longitudinal emittances
needed for a   high luminosity muon collider, 
\begin{equation}
 L = {{\gamma \, N^{\,2} f_0} \,\,(D\,C) \over {4 \pi  \epsilon_{x,y} \, \beta^{\,*}\rule{0pt}{9pt}}}  = {{7000 \, (2 \times 10^{\,12})^2 \ 180,000/{\rm{s}} \,\,(0.062)} \over {4 \pi \, (0.0025\,{\rm{cm}}) \, 1.0\,{\rm{cm}}}} 
 = {{1.0 \times 10^{\,\,34}}
 {\rm{\,cm}^{\,{-2}} \, s^{-1}}}
\end{equation}

\noindent
where $L$ is average luminosity, $N$ is the initial number of muons per bunch (one positive and one negative),
$f_0$ is the collision frequency (two detectors), 
$D\,C$ is the duty cycle with a 15\,Hz repetition rate, 
and $\beta^{\,*}$ is the betatron function in the collision region.
The initial 6D emittance must be small  enough, the potato slicer does not cool muons.
The longitudinal emittance is as large as can be tolerated by the $\sigma_p /p = 10^{\,-3}$ chromaticity requirement\,\cite{Alexahin}  of 
a 1.5~TeV/c$^{\,2}$ muon collider final focus  with round 750 GeV beams, $\beta$ = 0.99999999, 
$\gamma$~=~E/m$_{\mu}$, and a 10\,mm long collision region.

\begin{equation}
\epsilon_{L,N} = (\sigma_p /p) \, \Delta z \, (\beta \, \gamma) = 10^{\,-3} \, 10{\rm{mm}} \, 7000 = 70\,{\rm{mm}}
\end{equation}

\bigskip
\leftline{\large \bf Acknowledgements}
\medskip

Many thanks to Yuri Alexahin, Chuck Ankenbrandt, Scott Berg, Chandra Bhat, Alex Bogacz, Moses Chung,  Mary Anne Cummings, 
Jean-Pierre Delahaye,
Ben Freemire,
Carol Johnstone,
Trey Lyons,
Bob Palmer, 
Mark Palmer,
Philippe Piot,
Tom Roberts, Robert Ryne, Hisham Sayed, 
Pavel Snopok,
Diktys Stratakis, Mike Syphers, Yagmur Torun,
and Katsuya Yonehara for many useful conversations.

\bigskip
\leftline{\large \bf Appendix A: Skew quadrupole triplet algebra}
\medskip

\leftline{Define magnetic normalized emittance as} 
\smallskip

\leftline{$\epsilon_{mag,N} = \frac{e\rule[-7pt]{0pt}{7pt}B_{solenoid}\,\sigma^{2}_{(x,y)}}{2mc}$.}
\bigskip

\leftline{Define transverse intrinsic normalized emittance as} 
\smallskip

\leftline{$\epsilon_{TR,int,N} = \frac{\rule[-5pt]{0pt}{5pt}p_{beam}}{mc}\sigma_{(x,y)}\sigma_{(x',y')}$ 
which assumes $\sigma_{x} = \sigma_{y} = \sigma_{(x,y)}$ and $\sigma_{x'} = \sigma_{y'} = \sigma_{(x',y')}$.}
\bigskip

\leftline{When exiting solenoid, $(\epsilon_{x,N}, \epsilon_{y,N}) = (\epsilon_{TR,int,N},\epsilon_{TR,int,N})$ is transformed to $(\epsilon_{TR,smaller,N},\epsilon_{TR,larger,N})$}
\smallskip 

\leftline{with $\epsilon_{N,smaller} = \sqrt{\epsilon_{mag,N}^2+\epsilon_{TR,int,N}^2}-\epsilon_{mag,N} \approx \frac{\rule[-5pt]{0pt}{5pt}\epsilon_{TR,int,N}^2}{2\epsilon_{mag,N}}$ and}
\smallskip 

\leftline{$\epsilon_{N,larger} = \sqrt{\epsilon_{mag,N}^2+\epsilon_{TR,int,N}^2}+\epsilon_{mag,N} \approx 2\epsilon_{mag,N}$}
\bigskip

\leftline{Define $\beta_{\perp} = \frac{\rule[-5pt]{0pt}{5pt}2p_{beam}}{eB_{solenoid}}$.}
\smallskip

\leftline{$f_{Q1} = \frac  {\rule[-5pt]{0pt}{5pt}\beta_{\perp}}  {\rule{0pt}{19pt}\sqrt{1+\frac{\rule[-5pt]{0pt}{5pt}\beta^{2}_\perp}{L(L+M)}}}$}
\smallskip

\leftline{$f_{Q2} = \frac{\rule[-3pt]{0pt}{3pt}LM}{\beta_\perp}\left[1+\sqrt{1+\frac{\rule[-4pt]{0pt}{4pt}\beta^2_\perp}{L(L+M)}}\,\right]$}
\medskip

\leftline{$f_{Q3} = \frac{\rule[-4pt]{0pt}{4pt}2M(L+M)}{\beta_\perp}$}
\bigskip

\leftline{The relationship between the quadrupole focal length and integrated magnetic field is}
\smallskip

\leftline{$\frac{\rule[-4pt]{0pt}{4pt}p_{beam}}{ef_{quad}\rule{0pt}{8pt}} 
= \int\frac{\rule[-3pt]{0pt}{3pt}dB_y}{dx\rule{0pt}{8pt}}dl = \int\frac{\rule[-2pt]{0pt}{2pt}dB_x}{dt}dl \approx \frac{\rule[-4pt]{0pt}{4pt}B_{pole}}{r_{pole}}l_{quad}$}

\begin{figure}[b!]
\begin{center}
\epsfig{figure=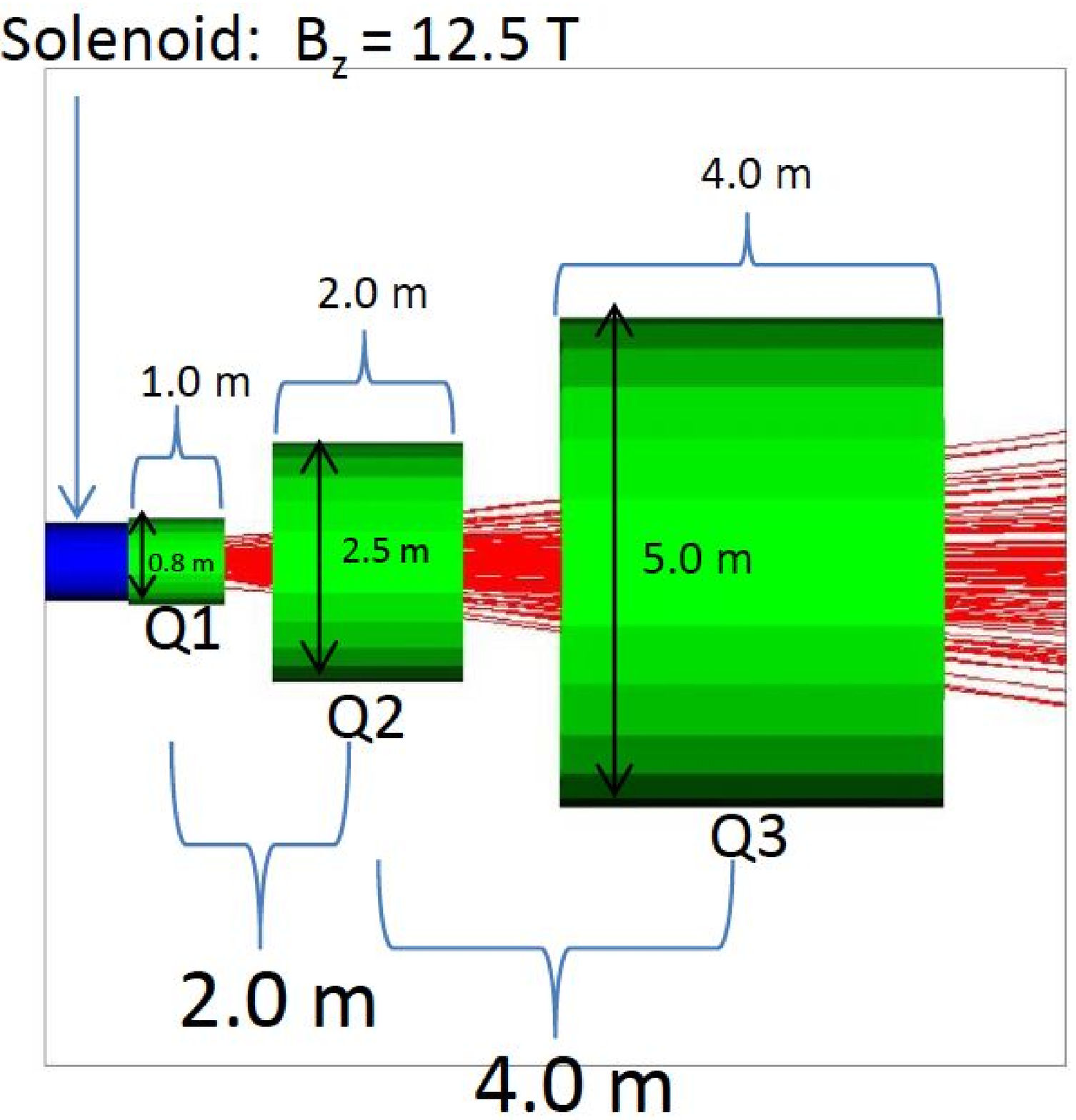,height=67mm}
\epsfig{figure=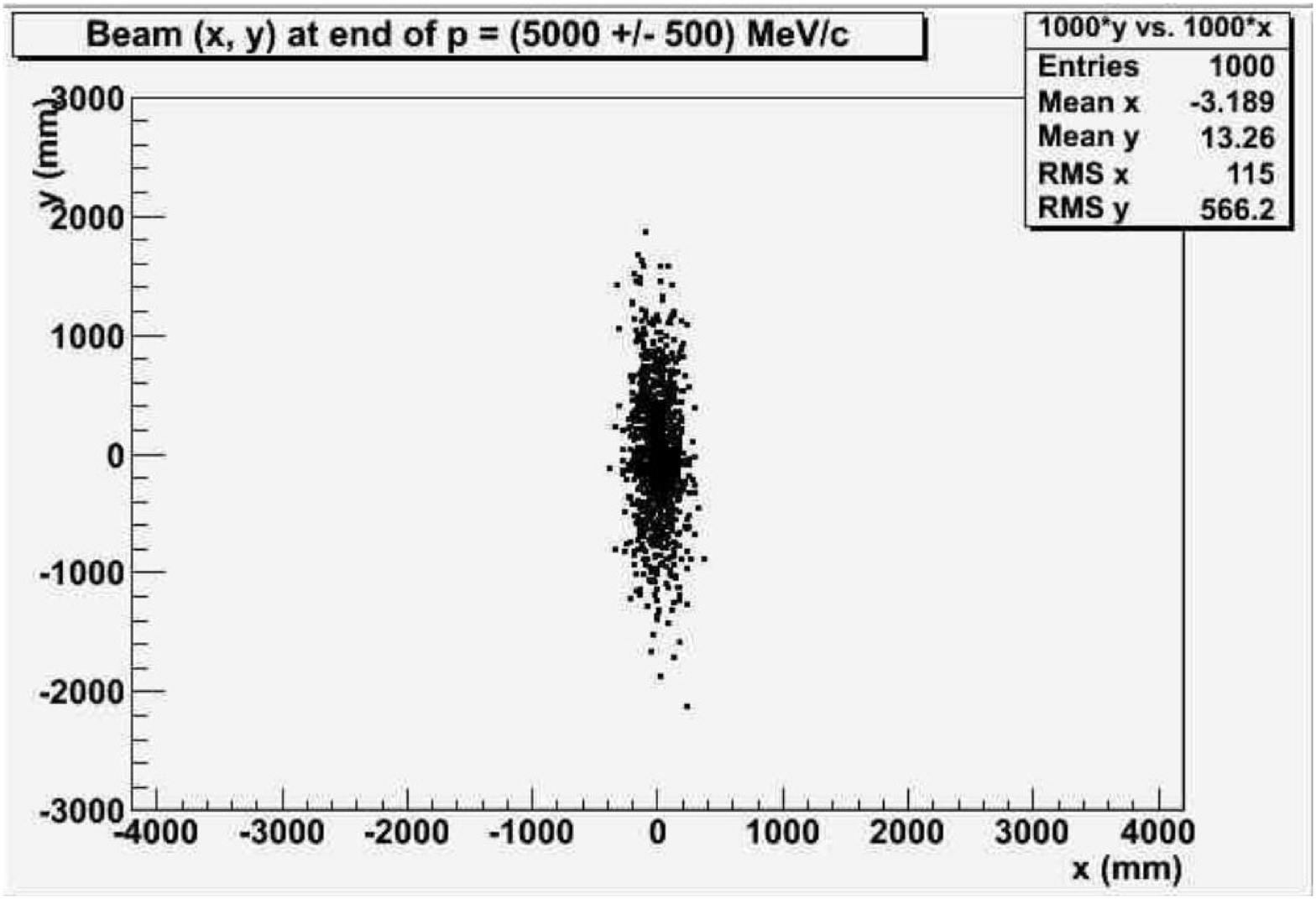,height=67mm}
\caption{A round, spinning 5000$\,\pm$500 MeV/c muon beam with a normalized transverse emittance of 50\,mm is transformed to a flat, non-spinning muon beam with new normalized
transverse emittances of 13\,mm and 295\,mm\,\cite{Brinkmann}. 
Inside the 12.5\,T solenoid $\sigma_{x,y}$~=~75\,mm.
The skew quadrupole pole tip fields 
are 
(4.49, 2.79, 0.684)m. 
The calculated thin quadrupole parameters must be adjusted somewhat  to make finite length quadrupoles work properly.
The simulation was
done with G4Beamline\,\cite{Roberts} for the solenoid and ICOOL\,\cite{Fernow2005} which works better for quadrupoles.}
\label{figure:Round2}
\end{center}
\end{figure}

\bigskip
\leftline{\large \bf Appendix B: Skew Quadrupole Triplet for a Large Emittances.}
\medskip
The emittance exchange of a large 50 mm transverse emittance beam is simulated.
Parameters are  chosen (see Fig.\,2)  to give an ideal emittance ratio of 16.  The actual  
new nominal transverse normalized emittances are  13 mm and 295 mm. Round to
flat beam transformations may have many uses.
A beam with a low emittance in one dimension might be spread with a dipole to separate isotopes\,\cite{Bertrand}.

\newpage
\leftline{\large \bf Appendix C: Deflecting RF Transverse to Longitudinal Exchange}
\medskip

\begin{figure}[t!]
\begin{center} 
  \epsfig{figure=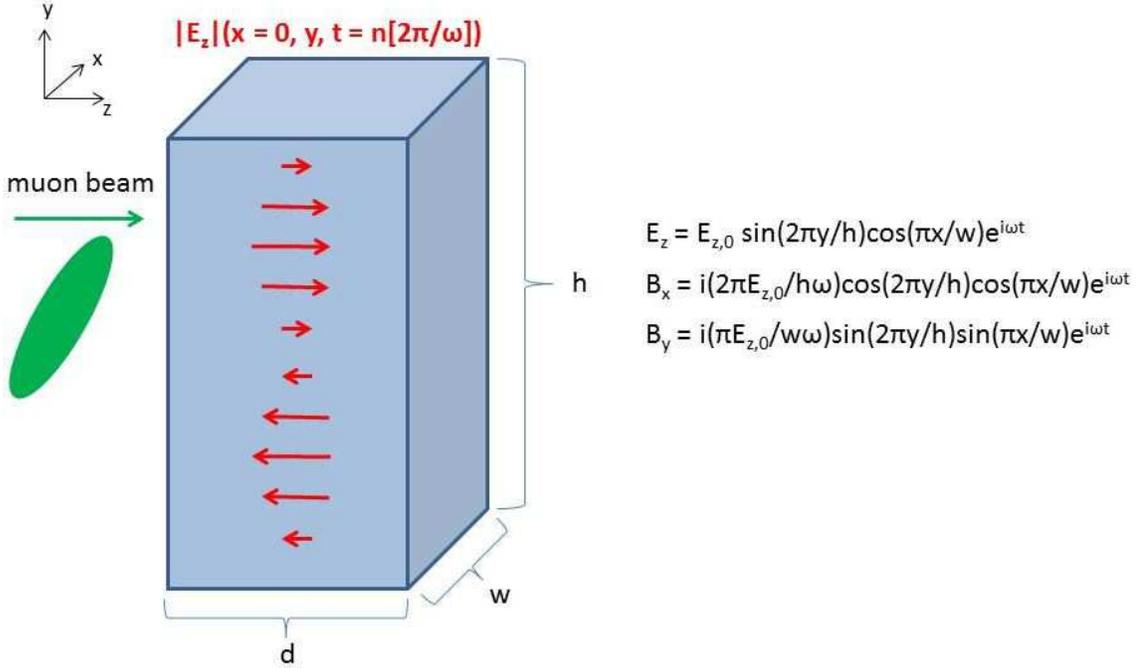,width=150mm}
  \caption{Schematic diagram of a box deflecting RF cavity with electromagnetic fields.}
  \label{fig:cavity}
\end{center}
\end{figure}

We explore the possibility of using deflecting RF cavities to transfer transverse emittance to longitudinal emittance.  The desired normalized emittance transfer 
is $(\epsilon_{y,N}, \epsilon_{L,N}) \sim (400, 1600)~\mu {\rm{m}} \rightarrow (25, 25600)~\mu$m so that ($\epsilon_{y,N}/\epsilon_{L,N}$) is (decreased/increased) 
by a factor of 16.  $\epsilon_{x,N}$ would be kept at 25 $\mu$m.  Figure~\ref{fig:cavity} shows a schematic diagram of a box RF deflecting cavity.
The inital muon momentum is roughly 200 $\pm$ 10 MeV/c, and the Gaussian beam position spread in $y$ is 30 mm.  Transforming the $y$ 
and longitundinal normalized emittances by a factor of 16 with deflecting RF cavities involves increasing the beam momentum spread 16-fold to 160 $MeV/c$ and 
establishing a correlation between $y$ and the momentum.  As this treatment is very approximate, the longitudinal quantity $z'$ will be defined as $\delta_p/p_0$ so 
that the following derivations may be missing  a factor or two of $\beta$ which is 0.884 for a 200 $MeV/c$ muon.   

\begin{figure}[t!]
\begin{center}
  \epsfig{figure=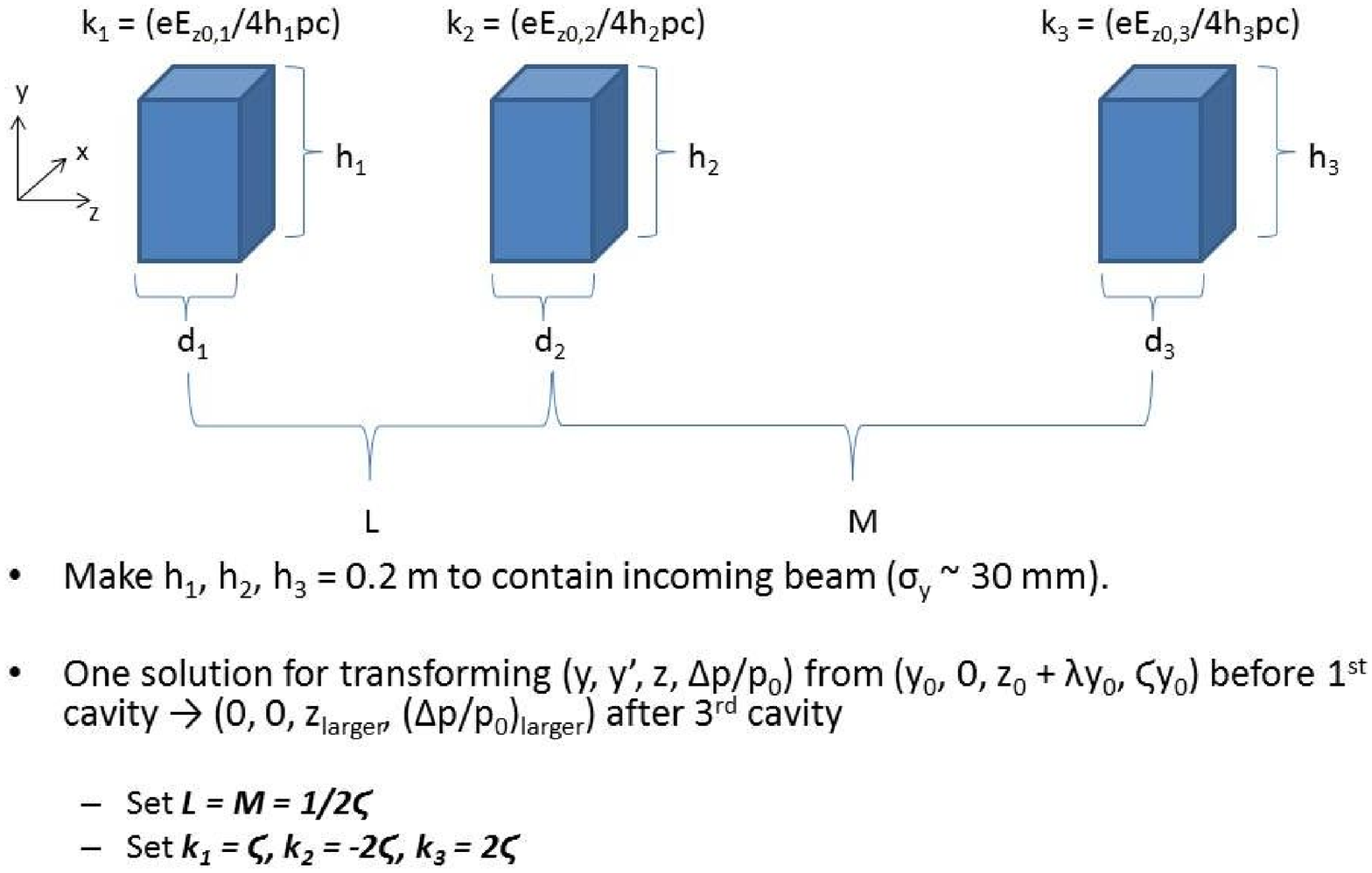,width=150mm}
  \caption{RF deflecting cavity triplet  that transforms transverse to longitudinal emittance.}
  \label{fig:cavity_layout}
  \end{center}
\end{figure}

An algorithm\,\cite{Ryne}  is used for determining the optics of a triplet of deflecting RF cavities that will enact the transvere-longitudinal emittance transformation.  Figure~\ref{fig:cavity_layout} shows the layout of a triplet of RF deflecting cavities.

There are an infinite number of solutions, and section VI\,-A of \cite{Ryne} includes a solution in which the three deflecting RF cavities are equally spaced and the 
normalized strength of the first deflecting RF cavity, $b$,  is set equal to the correlation ratio between the energy deviation and the transverse 
position, $\zeta \equiv z'/y_0 = (\delta_p/p_0)/y_0$.  
Defining the normalized strengths of the deflecting RF cavites as $k1$, $k2$, and $k3$ and the distances between the deflecting RF cavities as $L$:

$k1 = \zeta$ (first deflecting RF cavity)

$k2 = -2\zeta$ (second deflecting RF cavity)

$k3 = 2\zeta$ (third deflecting RF cavity)

$L = 1/2\zeta$ (distances between RF cavities)

\smallskip

Distances and strengths of the RF cavities will be now be worked out, which has $\zeta \sim (\delta_p/p_0)/y_0 \sim$ (160/200)/(0.03~m) = 26.7/m.  
Then,  L = distance~between~cavities = 1/(2 $\times$ 26.7/m) = 0.01875\,m.     
To capture the beam with $\sigma_y = 0.03~m$, the deflecting RF cavity should be about h = total height = 0.2\,m tall.  
The longitudinal electric field in a box deflecting RF cavity with total height $h$ (along $y$) and total width $w$ (along $x$) is
$E_z = E_{z,0}sin(2\pi y /h)cos(\pi x /w).$

\smallskip

Also, the total length, $d$, of deflecting RF cavites should be a fraction of the separation between the cavities, so that we can set $d$ = total cavity length = 0.01\,m.  Also, $d = \lambda_{RF}/2$ so that $\lambda_{RF}$ = 0.02\,m and  $f_{RF}$ = 15\,GHz.  Finally, the normalized strength, $k$ of a deflecting RF cavitiy is related to $E_{z,0}$ through $k = (e E_{z,0} d)/(4 h p_0 c)$ where $d$ and $h$ are the total length and total width of the deflecting RF cavity respectively.  For $d = 0.01~m$ and $h = 0.2~m$,

\bigskip
E$_{z,0}$ = 432,000~MV/m for~the~first~deflecting~RF~cavity,
\smallskip

E$_{z,0}$ = -864,000~MV/m for~the~second~deflecting~RF~cavity,
\smallskip

E$_{z,0}$ = 864,000~MV/m for~the~third~deflecting~RF~cavity.

\bigskip
The electric fields gradients required appear to be  vastly too high to perform a transverse to longitudinal exchange for large  emittances.

\newpage
\bigskip
\leftline{\large \bf Appendix Z: Quadrupole Doublet Focusing for Final Muon Cooling}
\medskip

\begin{figure}[b!]
\begin{center}
\epsfig{figure=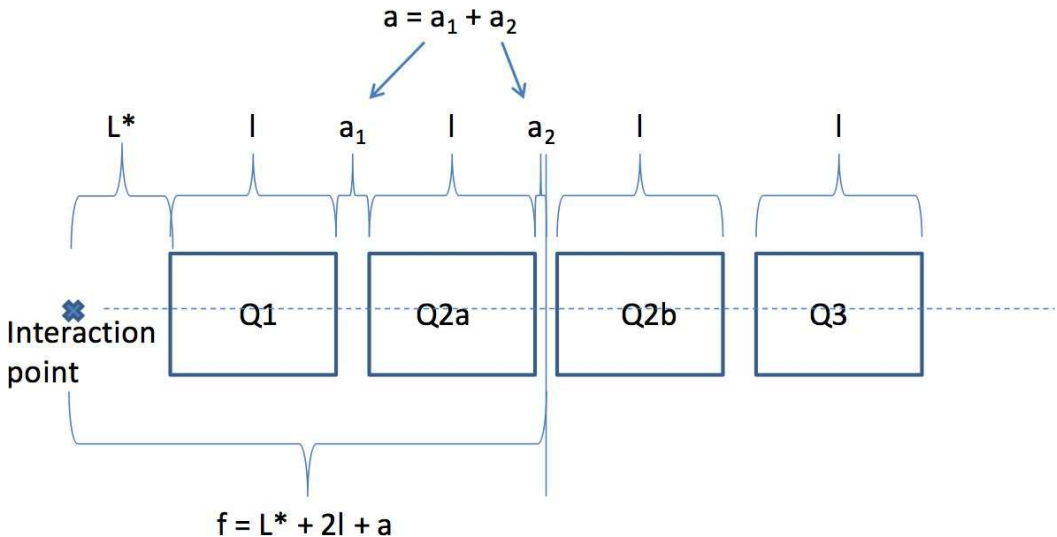,height=70mm} 
\vspace{-4mm}
\caption{Geometry of a quadrupole triplet focusing system\,\cite{Strait}.
Each quadrupole length is  $\ell$.
 L$^*$ is the distance from the interaction point to the front of the first quadrupole and $a$ is an additional length for magnet interconnections.
 $f = L_f = L^* + 2\,\ell + a$ is focal length. Magnet length times gradient is proportional to beam momentum divided by focal length.}
\label{figure:Quad3focus}
\end{center}
\end{figure}

\begin{figure}[t!]
\begin{center}
\epsfig{figure=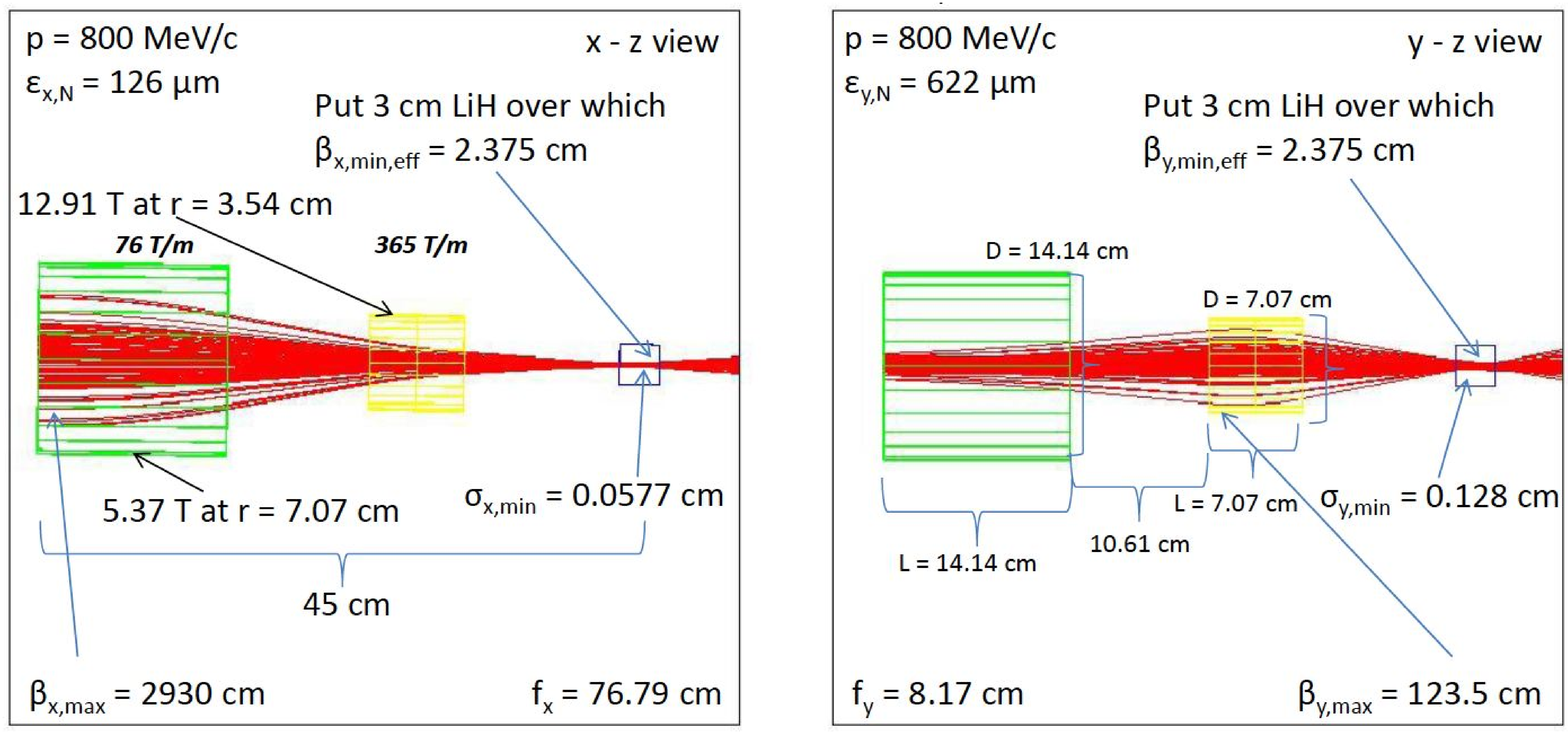,width=165mm} 
\caption{Quadrupole doublet half cell for final muon cooling with a flat beam, minimum $\beta_{x,y}$\,=~2\,cm, and a 3\,cm long LiH absorber.
The G4beamline transmission through one half cell is 998/1000 and the coverage for quadrupoles is at least $\pm  3.2 \sigma$.}
\label{figure:Quad2focus2cm}
\bigskip
\epsfig{figure=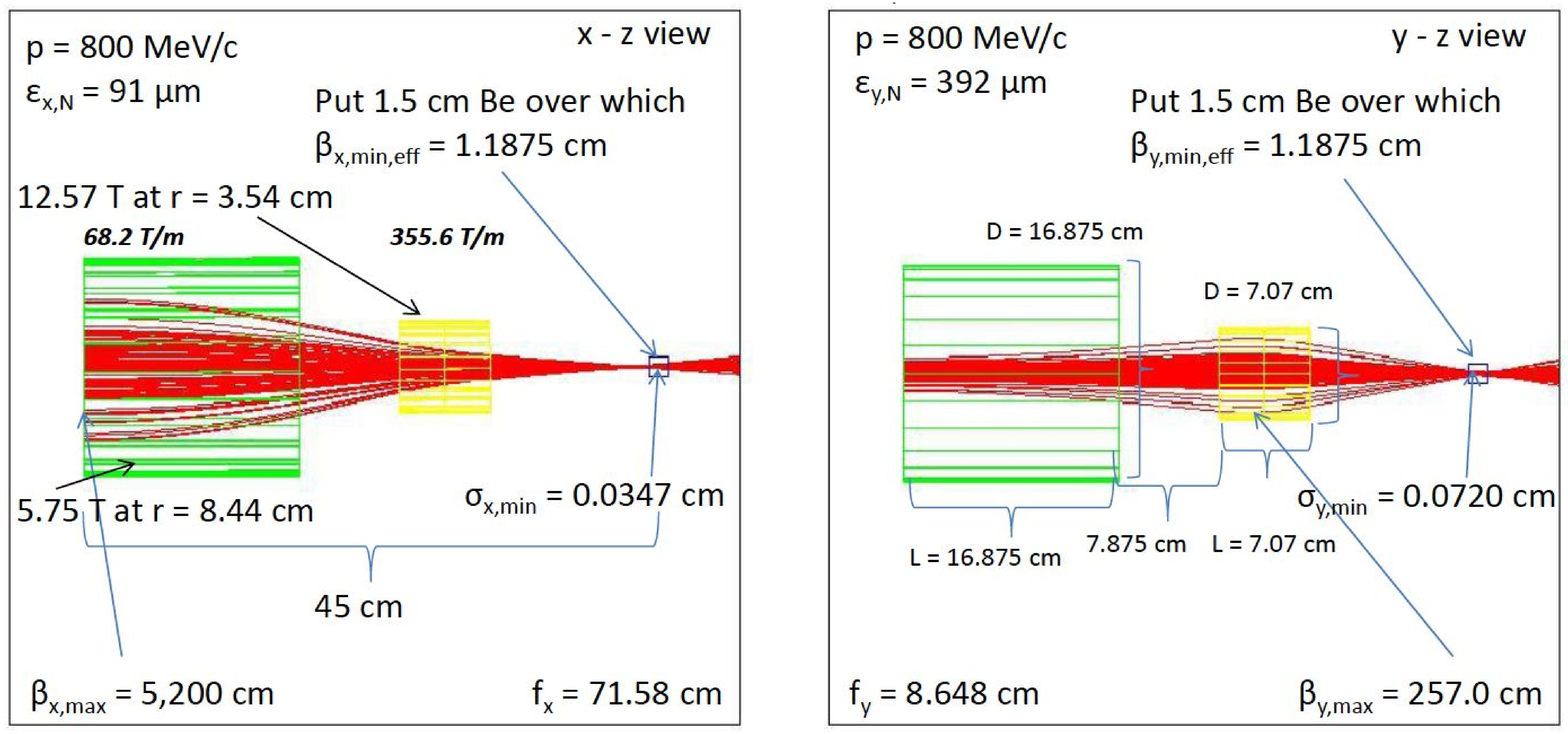,width=165mm} 
\caption{Quadrupole doublet half cell for final muon cooling with a flat beam, minimum $\beta_{x,y}$\,=~1\,cm, and a 1.5\,cm long beryllium absorber.
The G4beamline transmission through one half cell is 999/1000 and the coverage for quadrupoles is at least $\pm  3.06 \sigma$.}
\label{figure:Quad2focus1cm}
\end{center}
\end{figure}

Following Feher and Strait\,\cite{Strait} and their paper on hadron collider final focus quadrupole triplet design in 1996 with a $\beta^*$ of 50\,cm, we look into 
a  short quadrupole doublet for final muon cooling with a $\beta^*$ of 1\,cm. Focal lengths in x and y are not the same in the doublet.
The LHC inner triplet is composed of four identical 5.5\,m long quadrupoles as shown in Figure 5.
The outer two quadrupoles are focusing in the first  transverse dimension and defocusing
in the second transverse dimension. 
The inner two quadrupoles are focusing in the second  transverse dimension and defocusing
in the first transverse dimension. 
Add 0.3\,m for trim coils and take a quadrupole length, $\ell$, of 5.8\,m.
The focusing strength to be proportional to the lengths of the four quadrupoles 4$\ell$, times field gradient G in T/m, times focal length (L$_f$ = L$^*$ + 2$\ell$ + $a$) in meters,
divided by beam momentum $p$ in TeV/c. L$^*$ is the distance from the  interaction point to the front of the first quadrupole and $a$ is additional length for magnet
interconnections. The focal length is just the distance from the interaction point to the center of the quadrupole triplet as shown in Figures 6 and 7.
The relation between $\beta$ functions and the focal length is given by
$\beta_{\rm{\,max}} = b\,  L_f^2 / \beta^*$, where b is a fudge factor equal to 1.65 for the LHC.

A short length of  low Z absorber absorber is placed at the focus of each quadrupole doublet as shown in Figures 6 and 7.
Flat beams are used with the $\sin{(2\theta)}$  quadrupole doublets which do not exceed 14\,T as in the LHC Nb$_3$Sn LARP quadrupoles\,\cite{LARP}.
Note that $\beta(s) = \beta^* + s^2/\beta^*$.
As $\beta^*$ becomes smaller, the absorber
must   become thinner in the beam direction $s$, so one may want to employ beryllium or diamond.
The fringe fields of the magnet fall off as the cube of distance\,\cite{Johnstone}  and
may be small enough to not cause breakdown even in vacuum RF. 
The beam power of $4 \times 10^{12}$ 800 MeV/c muons (KE = 701 MeV) arriving at 15 Hz is 6700 watts of which only
a tiny fraction would heat the superconductor even in the absence of shielding.

In summary, quadrupole doublets and dense, low Z absorbers are being examined to cool the current outputs given either by the Helical\,\cite{Yoshikawa} or Rectilinear\,\cite{Stratakis} 
6D muon cooling channels and prepare the
input for the potato slicer which reduces the normalized transverse beam emittance to  the 25\,$\mu$m size required by a high luminosity muon collider. 
A more complete simulation of a tapered quadrupole doublet channel is in progress.

\bigskip

\end{document}